\documentstyle[12pt]{article}

\begin{document}
\input epsf %
\def\bmat#1{\left(\matrix {#1}\right)}
\begin{titlepage}
\centerline{\large \bf Generalized oscillator 
strength for Na $3s-3p$ transition}
\vglue .1truein
\centerline{Zhifan Chen and Alfred Z. Msezane}
\centerline{\it Center for Theoretical Studies of Physical Systems,  and 
Department of Physics}
\centerline{\it Clark Atlanta University, Atlanta, Georgia 30314, U. S. A.}
\centerline{\bf ABSTRACT}
Generalized oscillator strengths (GOS's) for
the Na $3s -3p$ transition have been investigated  
using the spin-polarized technique of the random phase
approximation with exchange (RPAE) and the first Born approximation (FBA), 
focussing our attention on the position of the minimum.
Intershell correlations are found to influence the position
of the minimum significantly, but hardly that of the maximum.
The RPAE calculation predicts for the first time the positions of
the minimum and maximum  
at momentum transfer, $K$ values of 1.258 a.u. and 1.61 a.u.,
respectively. The former value is within the
range of values extracted from experimental measurements, $K=1.0-1.67$
a.u.. We recommend careful experimental search for the minimum
around the predicted value for confirmation.

\centerline{}
PACS number(s):34.10.+x, 34.50Fa, 31.50.+w  
\end{titlepage}

\section{Introduction}
\ \ \ The generalized oscillator strength(GOS) 
is an important property of the atom, since Bethe introduced it [1]. 
To study this property,
the sodium atom has been chosen as the subject in many
experimental and theoretical investigations because
its electronic configuration has an inert core and a single
valence electron which is similar to that of the hydrogen atom. 
The differential cross sections (DCS's) and 
GOS's for the Na $3s-3p$ transition  
were measured by Shuttleworth {\it at al} [2] 
using a high-resolution electron spectrometer 
over the angular range of 1$^\circ$-20$^\circ$ 
at the incident electron energies 
of 54.4, 100, 150 and 250 eV. The measurement observed a GOS minimum  
at the momentum transfer value of $K=0.67$ a.u. (or $K^2=0.45$
a.u.). 
Buckman and Teubner [3] measured data for the same transition  
using a modulated crossed-beam technique 
at the incident energies of 54.4 , 
100 , 150  and 217.7 eV,   
covering the extended angular range of 2$^\circ$ to 145$^\circ$. 
Their data are in good agreement with those of Shuttleworth {\it et al}
at small angles but 
did not show a GOS minimum 
around $K^2=0.45$ a.u.. 

The Shuttleworth's {\it et al} minimum was also not confirmed
by the experiment of Srivastava and Vuskovic [4], 
which utilized a crossed-electron
-beam-metal-atom-beam scattering technique and performed measurements
at incident energies of 10, 20, 40, and 54.4 
eV.
The data
of Srivastava and Vuskovic are in poor agreement with 
those of Buckman and Teubner at large angles.
Srivastava and Vuskovic implied that the conflict was caused by
the inproper geometrical correction factor 
in Buckman and Teubner's
experiment.
To resolve the discrepancy Teubner {\it et al} [5] remeasured
the Na $3s-3p$ transition at
22.1 and 54.4 eV and 
found a probable source of systematic error in the measurement
of Srivastava and Vuskovic. All the measurements did not 
observe the GOS minimum around $K=0.67$ a.u. as predicted by Shuttleworth
{\it et al}.
Some recent measurements on the Na $3s-3p$ transition 
by Bielschowsky {\it et al} [6] 
at impact energy of 1 keV and  Marinkovic {\it et al} [7] at 10, 20 
and 54.4 eV
did not report
GOS minima in the momentum transfer regions they considered.

Theoretically,
Shimamura [8] predicted GOS minima to appear between $K^2=0.72$ 
and 0.93 a.u.,
depending on the choice of the exponent in the Slater orbitals.
Miller [9] calculated the GOS minimum in the Na $3s-3p$ transition
within the FBA, 
employing hydrogenlike orbitals with
effective nuclear charge and predicted a minimum 
at $K^2=0.71$ a.u..  
The FBA calculation from Bielschowsky {\it et al} 
showed a minimum around $K^2=2.0$ a.u..  
In summary the GOS minimum for the Na $3s-3p$ transition 
observed in Shuttleworth's {\it et al}
experiment was not confirmed by other measurements. 
Also, the wide range of positions of the minimum predicted 
by the theoretical calculations
at $K^2=0.71$, 0.72, 0.93 and 2.0 a.u. 
are not observed by the experments. Obviously, the GOS and 
the position of its minimum for the Na $3s-3p$ transition is still
an unresolved and interesting problem. 

In this paper
we have used the spin-polarized technique of the 
random phase approximation with exchange(RPAE) 
to investigate the GOS for the Na $3s-3p$ transition. The major
objective of our calculation has been the unambiguous identification
and location of the positions of the 
minimum and maximum. As a result, values of $K$ were carefully selected.
We found
for the first time the positions of the minimum and the maximum to be at
around $K=1.258$ a.u. (or $K^2=1.582$ a.u.) and $K=1.61$ a.u.
(or $K^2=2.59$ a.u.), respectively. 
After careful analysis of the experimental data, we found the 
GOS minimum at $K=1.258$ a.u. is supported by some previous
measurements. The intershell correlations are found to have
significant effect to the position of the minimum.   

\section{THEORY} 
In the FBA the generalized oscillator 
strength, f, for dipole allowed transitions in the length form 
can be calculated [10] as
\begin{equation}
f={2(2l+1)N_l w \over (2l_i+1)K^2}|d_\alpha|^2
\end{equation}
where $N_l$ is the number of electrons in the excited state, 
$l_i$ is the initial orbital angular momentum of the excited
electron,
$l$ is the total angular momentum of the electron-hole pair,
which satisfies triangle rule $|l_f+l_i|>l>|l_i-l_f|$,
$w$ is the excitation energy (a.u.). The  
dipole matrix element, $d_\alpha$ can be calculated from 
\begin{equation}
<\phi_f|d_\alpha|\phi_i>=\sqrt{(2l_i+1)(2l_f+1)}\bmat{l_f\ l\ l_i \cr 0\ 0\ 0 \cr}
\int_0^\infty P_i(r)P_f(r)
j_l(Kr)dr 
\end{equation} 
where $P_i(r), P_f(r)$ are the radial wave functions of the initial
and final states, respectively, $j_l(Kr)$ is the spherical
Bessel function. 
For the dipole allowed transition, the calculations performed
in this paper are with $l=1$ and $P_i(r), P_f(r)$ represented
by Hartree-Fock wave functions. Each channel, such as $s-p$
transition, includes three discrete excited states, $3p, 4p, 5p$,
and seventeen continuum states. The radial part of the wave function 
for each state was represented by 700 points.

According to the
semiempirical Hund rule, the total spin of a shell in the
ground state reaches the largest value permitted by the 
Pauli principle. Therefore, in the semiclosed shell all the 
electron spin vectors are collinear, and their projections on
to an arbitrary fixed direction are equal. Therefore,
every shell can be devided into two spin subshells,
each having a certain spin direction, $\uparrow$ or $\downarrow$. 
Because of this the wave functions double. Electrons in 
the subshell interact with
other electrons in two different ways, with or without exchange. 
Since the Coulomb interaction will not change the spin direction,
only the electrons 
having the same spin direction can interact with exchange. 

The equation for the dipole matrix element in the spin-polarized
technique of the RPAE [11] is  
\begin{equation}
(D^{\uparrow}_{\alpha},D^{\downarrow}_{\alpha})=(d^{\uparrow}_{\alpha},
d^{\downarrow}_{\alpha})+\sum_{\alpha'} (D^{\uparrow}_{\alpha'}
\chi^{\uparrow}_{\alpha'},D^{\downarrow}_{\alpha'}\chi^{\downarrow}_{\alpha'})
\bmat{u^{\uparrow\uparrow}_{\alpha'\alpha}u^{\uparrow\downarrow}
_{\alpha'\alpha} \cr
u^{\downarrow \uparrow}_{\alpha'\alpha} u^{\downarrow \downarrow}
_{\alpha'\alpha} \cr}
\end{equation}
where $u_{\alpha' \alpha}$ is the Coulomb inter-electron potential, 
$d^{\uparrow}_\alpha$ represents the amplitude for the direct excitation of 
$3s^\uparrow - 3p^\uparrow$ 
and $\chi_{\alpha'}$ is the
electron-vacancy propagator. 
If the states of $\uparrow$ and $\downarrow$ electron are equivalent,
the dipole matrix element $D^\uparrow_{\alpha}$
in the RPAE for the $3s^\uparrow - 3p^\uparrow$ transition 
can be obtained from 
\begin{eqnarray}
<\epsilon_f|D^{\uparrow}_{\alpha}|\epsilon_i>&=&<\epsilon_f|d^
{\uparrow}_{\alpha}
|\epsilon_i>+(\sum_{\epsilon_3\le F,\epsilon_4 >F} -
\sum_{\epsilon_3>F, \epsilon_4\le F}) \nonumber   \\  
&&{<\epsilon_4|D^{\uparrow}_{\alpha} 
|\epsilon_3><\epsilon_3 \epsilon_f|u^{\uparrow \uparrow,\uparrow \downarrow},
|\epsilon_4 \epsilon_i>
\over w-\epsilon_4 +\epsilon_3 +i\eta(1-2 n_4)}   
\end{eqnarray}
where $\epsilon_3$ and $\epsilon_4$ represent the virtual excitation states, 
$i \eta$ gives the direction of tracing the pole while integrating over
the energy, $\eta \rightarrow +0$, $F$ is the Fermi energy of the atom, 
and $n_4$ is the Fermi step: $n_4=1$, $\epsilon_4 \le F$; $n_4=0$,
$ \epsilon_4 >F$.
The symbol $\sum$ denotes summation over discrete and integration 
over continuous states. A similar equation for the $D^\downarrow_\alpha$
of the $3s^\downarrow - 3p^\downarrow$ transition can be obtained.
It is important to remember that  
only the states with the same spin direction
can have exchange interaction in the
sum of Eq. (4).
Finally, the GOS for Na $3s-3p$ transition can be
written as
\begin{equation}
f={2(2l+1)N_l w \over (2l_i+1)K^2}
({D^{\uparrow}_{\alpha}} ^2+{D^{\downarrow}_{\alpha}} ^2)
\end{equation}
Eqs. (1),(2),(4) and (5) are the basic equations used in this paper to 
calculate the GOS's in FBA 
and RPAE.

\section{RESULTS AND DISCUSSION}
The results of our calculation are given in Fig. 1, Fig. 2 and 
Table 1.  Table 1 lists the positions of the minimum and maximum for
the Na $3s-3p$ transition obtained by different authors. The results
of our FBA calculation are in excellent agreement
with that of Bielschowsky {\it et al}. The data of 
Shuttleworth {\it et al} showed a GOS minimum around $K=0.67$ a.u..
However, this minimum was not observed in all other measurements. 
After analyzing the experimental data obtained by Shuttleworth
{\it et al} [2], Buckman and 
Teubner [3], Srivastava and Vuskovic [4], Teubner {\it et al} [5], and
Marinkovic {\it et al} [7], we found that
the GOS minima from the experimental data are around $K=1.0-1.67$
a.u. (or $K^2=1.0-2.80$ a.u.)
at impact energies of 20 and 54.4 eV. These minima were not originally
noticed by the experimentlists. 

\epsfbox{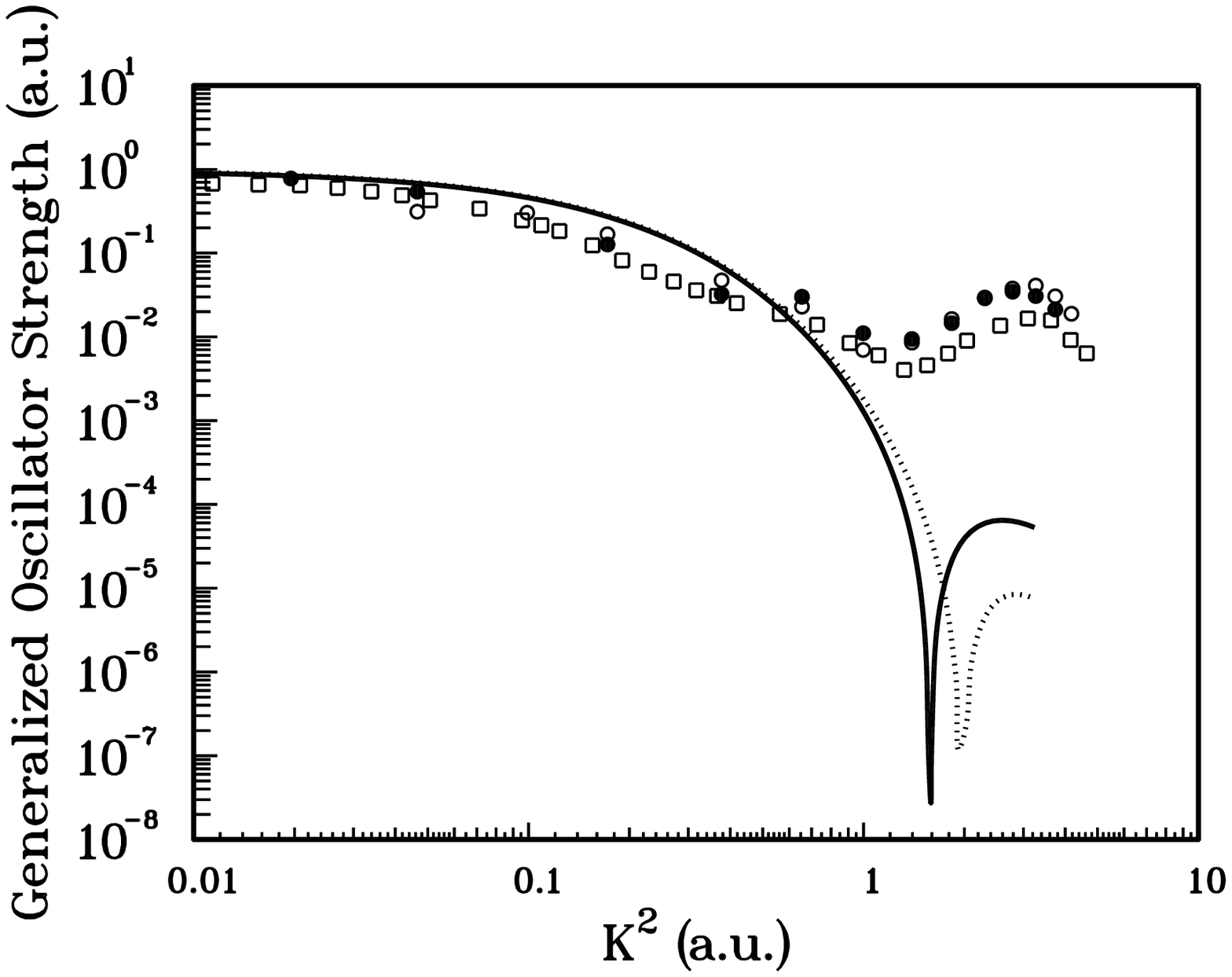}

Fig. 1 shows the GOS's versus $K^2$ for the Na $3s-3p$ transition
at 20 eV. The circles are from Srivastava and Vuskovic, black dots are
from Marinkovic {\it et al}, and squares are from Teubner {\it et al}
. The solid and dotted lines represent our RPAE and FBA calculations,
respectively. All three measurements show the GOS minimum around
$K=1.0-1.18$ a.u. (or $K^2=1.0-1.39$ a.u.) which are close to our results.

\epsfbox{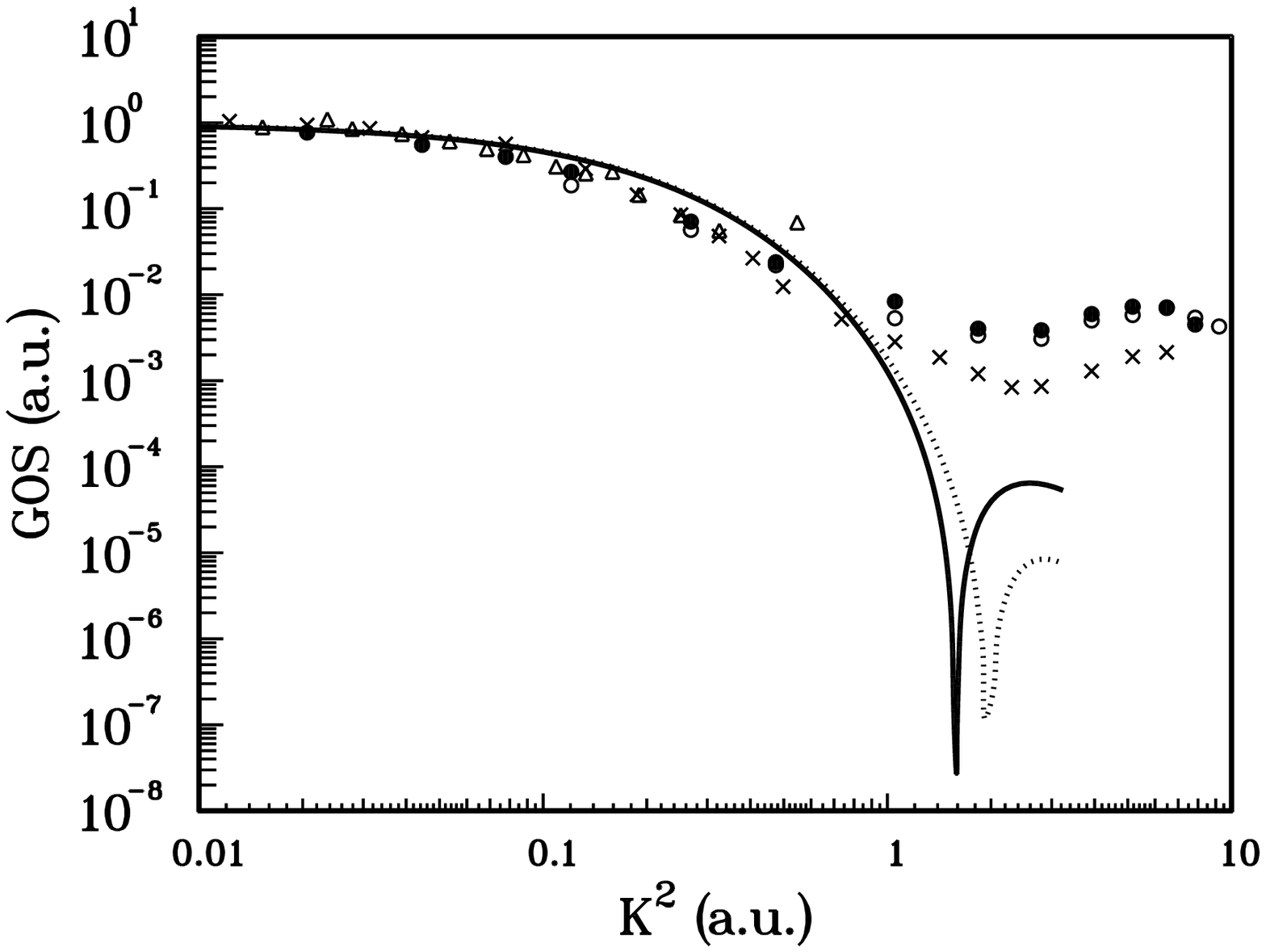}

Fig. 2 is the same as Fig.1, except that the data are 
at 54.4 eV and the crosses are from
Buckman and Teubner and the triangles are from Shuttleworth {\it et al}.
The black dots and circles show the GOS minimum around $K=1.67$ a.u.
(or $K^2=2.80$ a.u.),
while the crosses give a minimum at $K=1.51$ a.u. (or $K^2=2.29$ a.u.).
Since the position of the GOS minimum for the Na $3s-3p$ transition
occurs at large momentum transfer, $K=1.0-1.67$ a.u., which
corresponds to large angles in the measurements; for  
example $\theta=37^\circ$ at the impact energy of 54.4 eV($K=1.258$ a.u.).
Therefore it is easy to miss the GOS minimum in the measurement by taking
large angular steps at large angles as is the usuall practice. 
From our calculation
it is suggested to reperform the experiment for Na $3s-3p$ transition
and pay particular attention to the position of the minimum 
we indicated in Table 1. We believe more experimental data will
be obtained to confirm our results. 

It is interesting  
to compare the intershell correlations for the transitions between
the Na $3s-3p$ and the Ar $3p-4s$. The position of the GOS minimum for
the later is influenced insignificantly by correlations.
The difference between positions of the minimum 
from the RPAE and FBA is less than 0.7\%.
However, in the Na $3s-3p$ transition the difference is more
than 11\%.  This is because the Na $3p$ level is only 2.1 eV above the $3s$ ground
state; therefore Na has an enormous dipole polarizability 
$(23.6 \times 10^{-24}cm^3)$.
In RPAE the many-electron correlation effects are essentially
due to polarization of the electron shell by the external field.
Results from RPAE will show a significant difference from that of FBA
if the atoms have large polarizability. Therefore, we can expect 
that Li, K and Rb
atoms will yield results that are similar to those of Na,
while Ne, Kr and Xe atoms will exhibit results similar
to those of Ar if we compare the positions of GOS minimum from RPAE and FBA. 

Correlations between intershell electrons are strongly affected by the
separation of the subshells. The $3s\uparrow-3p\uparrow$ excitation energy will be  
0.07267(a.u.), 0.07255(a.u.), 0.07190(a.u.), and 0.07175 (a.u.)
if channels ($s^\uparrow-p^\uparrow$), ($s^\uparrow-p^\uparrow
+p^\uparrow-s^\uparrow+p^\downarrow-s^\downarrow$), 
($s^\uparrow-p^\uparrow+p^\uparrow-d^\uparrow+p^\downarrow-d^\downarrow$)
and ($s^\uparrow-p^\uparrow+p^\uparrow-d^\uparrow+p^\downarrow-d^\downarrow
+p^\uparrow-s^\uparrow+p^\downarrow-s^\downarrow$) are included 
in the RPAE calculations.  This indicates that the influence of the 
$3d^{\uparrow
\downarrow}$ electrons
upon the 
$3p\uparrow$ electron is larger than that from the 
$4s^{\uparrow \downarrow}$ electrons.
 
In conclusion, the positions of the minimum and the maximum for the 
Na $3s-3p$
transition have been calculated and found for the first time from
RPAE at the 
momentum transfer
values of $K=1.258$ a.u. and 1.61 a.u., respectively. Furthermore, the
many-electron correlations are found to 
play an important role in the determination of the position of the minimum. 
We recommend that experiments search carefully for predicted minimum,
using the value obtain here as a guide.

\section{Acknowledgments}
Research was supported in part by the US DoE, Division of Chemical Sciences, Office of Basic
Energy Sciences, Office of Energy Research and NSF.

\newpage 

\begin{table}
\caption{Positions of the GOS minimum and the maximum for 
the excitation of Na $3s-3p$}   
\begin{center}
\vskip 0.1 truein
\begin{tabular}{cccccc}
\hline
\hline
\multicolumn{4} {r} {$K$ (a.u.) for minimum}
&\multicolumn{2} {c} {$K$ (a.u.) for maximum}\\ 
Atom& Authors  & Expt.         & Theory  & Theory \\  \hline  
    & Shuttleworth {\it et al} [2]        & 0.67   &    &     \\    
    & Marinkovic {\it et al} [7]          & 1.18-1.67 & & \\
    & Buckman and Teubner [3]             & 1.51        && \\
    & Srivastava and Vuskovic [4]          &1.0-1.7  &&\\
    & Teubner {\it et al} [5]             &1.15        &&\\
    & Shimamura                           &        & 0.847-0.965 &    \\
Na  & Miller                              &        & 0.84 &         \\
    & Bielschowsky {\it et al}            &        & 1.41 &    \\ 
    & Present RPAE                        &        & 1.258  &1.61  \\
    & Present H-F                         &        & 1.401  & 1.69\\
\hline 
\end{tabular}
\end{center}
\end{table}

\newpage
\leftline{\bf Figure Captions}
Fig. 1. GOS's versus $K^2$ for the Na $3s-3p$ transition at 20 eV.  
Circles are from Srivastava and Vuskovic, black circles are from 
Marinkovic {\it et al}, and squares are from Teubner {\it et al}.
Solid and dotted lines represent   
our RPAE and FBA calculations, respectively. 

Fig. 2. GOS's versus $K^2$ for the Na $3s-3p$ transition at 54.4 eV.
All the symbols have the same meaning 
as in Fig. 1, except that 
crosses are from Buckman and Teubner and triangles are from 
Shuttleworth {\it et al}. 
\end{document}